\documentclass[conference]{IEEEtran}
\IEEEoverridecommandlockouts
\usepackage{cite}
\usepackage{amsmath,amssymb,amsfonts}
\usepackage{algorithmic}
\usepackage{graphicx}
\usepackage{textcomp}
\usepackage{xcolor}
\usepackage{multirow}
\usepackage{subfig}
\usepackage{booktabs} 
\def\BibTeX{{\rm B\kern-.05em{\sc i\kern-.025em b}\kern-.08em
    T\kern-.1667em\lower.7ex\hbox{E}\kern-.125emX}}
\begin{document}

\title{Towards Multi-Task Generative-AI Edge Services with an Attention-based Diffusion DRL
Approach}

% \title{Conference Paper Title*\\

\author{
    \IEEEauthorblockN{\large
    Yaju Liu\IEEEauthorrefmark{1}\IEEEauthorrefmark{2},
    Xi Lin\IEEEauthorrefmark{1}\IEEEauthorrefmark{2},
    Siyuan Li\IEEEauthorrefmark{1}\IEEEauthorrefmark{2}, 
    Gaolei Li\IEEEauthorrefmark{1}\IEEEauthorrefmark{2},
    Qinghua Mao\IEEEauthorrefmark{1}\IEEEauthorrefmark{2},
    Jianhua Li\IEEEauthorrefmark{1}\IEEEauthorrefmark{2}\\}
    \IEEEauthorblockA{\IEEEauthorrefmark{1}\textit{School of Electronic Information and Electrical Engineering, Shanghai Jiao Tong University, Shanghai, China} \\}
    \IEEEauthorblockA{\IEEEauthorrefmark{2}\textit{Shanghai Key Laboratory of Integrated Administration Technologies for Information Security, Shanghai, China} \\}
    % \thanks{Corresponding author: Xi Lin (linxi234@sjtu.edu.cn).}
}

\maketitle
% Selecting the suitable AIGC service according to the type of the AIGC task and the left computing resources on the edge server.
\begin{abstract} 
As an emerging paradigm of content creation, AI-Generated Content (AIGC) has been widely adopted by a large number of edge end users.
However, the requests for generated content from AIGC users have obvious diversity, and there remains a notable lack of research addressing the variance in user demands for AIGC services.
This gap underscores a critical need for suitable AIGC service selection mechanisms satisfying various AIGC user requirements under resource-constrained edge environments.
% the same AIGC model applied to different types of user tasks.
% the edge application of various AIGC models is constrained by limited resources in edge environments.
To address this challenge, this paper proposes a novel Attention-based Diffusion Soft Actor-Critic (ADSAC) algorithm to select the appropriate AIGC model in response to heterogeneous AIGC user requests. 
Specifically, the ADSAC algorithm integrates a diffusion model as the policy network in the off-policy reinforcement learning (RL) framework, to capture the intricate relationships between the characteristics of AIGC tasks and the integrated edge network states.
Furthermore, an attention mechanism is utilized to harness the contextual long-range dependencies present in state feature vectors, enhancing the decision-making process.
Extensive experiments validate the effectiveness of our algorithm in enhancing the overall user utility and reducing servers crash rate.
Compared to the existing methods, the proposed ADSAC algorithm outperforms existing methods, reducing the overall user utility loss and the server crash rate by at least 58.3\% and 58.4\%, respectively. 
These results demonstrate our ADSAC algorithm is a robust solution to the challenges of diverse and dynamic user requirements in edge-based AIGC application environments.
\end{abstract}


\section{Introduction}
AIGC represents a novel paradigm in content generation using Generative Artificial Intelligence (GAI) models and has ushered in a new era in which the automated generation of diverse high-quality content in a short time becomes possible. With the introduction of transformers and diffusion models in the content generation domain in recent years, the demonstrated strong natural language processing and high-quality image generation capabilities have driven the rapid development of AIGC services, such as ChatGPT \cite{ChatGPT} and Stable Diffusion \cite{stablediffusion}. The capabilities of AIGC extend to a wide range of generated content forms, including text, image, audio, and video. Meanwhile, AIGC has attracted considerable interest across various sectors, demonstrating its potential as a transformative tool in healthcare \cite{chen2024revolution}, medicine \cite{shao2024artificial}, education \cite{zhu2023chatgpt}, marketing \cite{lin2023research}, and manufacturing \cite{wang2023industrial}.

% 边缘网络赋能AIGC的广泛使用
Leveraging the rapid advances in GAI technology, the quality and diversity of generated content are significantly enhanced, allowing AIGC to be increasingly integrated into existing digital platforms, tools, and services. Despite the significant advances in hardware technology for portable mobile devices such as smartphones, handheld devices, and wearable devices, the computational intensity of GAI models remains a bottleneck for the widespread application of AIGC services \cite{luong2019applications}. While some mobile devices can support the inference of GAI models, this often results in considerable processing latency and non-negligible battery energy consumption \cite{ho2020joint}. Therefore, the direct use of AIGC applications on mobile devices is challenging. Fortunately, the development of edge computing and the advances in wireless network technology are expected to mitigate these computational constraints and enable more devices to access and use AIGC services efficiently \cite{du2024diffusion, xu2024unleashing}. Using high-bandwidth and low-latency wireless networks, user devices offload prompts and relevant requirements for content generation tasks to edge servers. These edge servers, equipped with GAI models, perform model inference and generate the appropriate content based on user prompts and requirements. The content is then transmitted back to the respective devices via the wireless network. Benefiting from the advances in 5G and upcoming 6G technologies \cite{celik2024dawn}, AIGC services will be deployed in various scenarios more widely and efficiently.

\begin{figure}[!t]
    \centering
    \includegraphics[width=\linewidth]{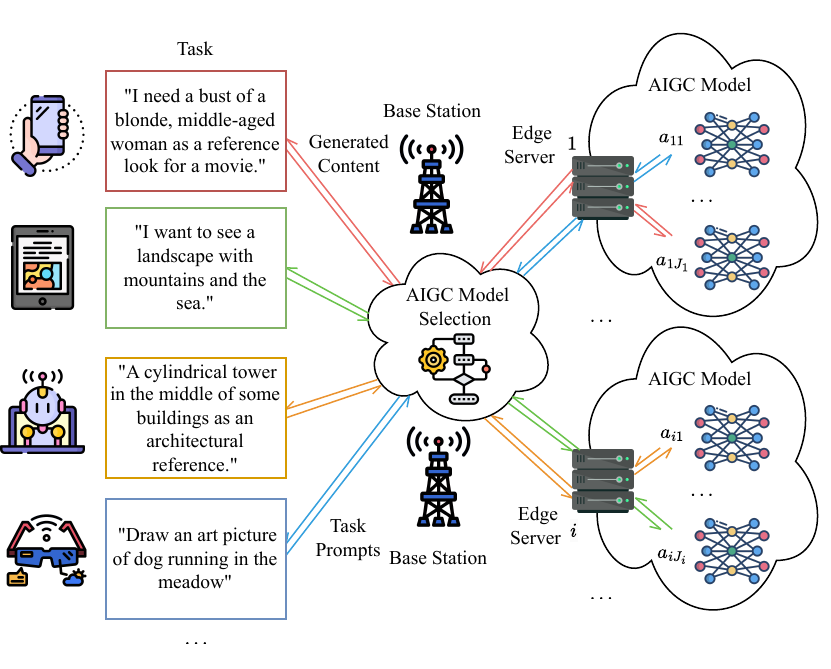}
    \caption{Edge computing-supported AIGC service architecture for multi-type task scenario.}
    \label{fig1}
\end{figure}

% 现有方法，提出用户任务种类多样的场景
Various AIGC models have been published to satisfy the demands of users. Considering the variation among AIGC models and resource constraints, an AIGC model selection algorithm is needed to select the most appropriate AIGC model. Besides, user tasks have multiple types, such as text-to-image generation tasks including portraits, landscapes, architecture, and art paintings, as shown in Fig.~\ref{fig1}. The generative capability of an AIGC model varies for different task types, and different AIGC models offer different levels of user utility for the same task type. Therefore, the AIGC model selection algorithm must also consider the specific nature of user tasks as a critical factor in the decision. 

% 方法介绍
This paper explores the problem of AIGC model selection in multi-type user task scenarios under resource constraints to maximize overall user utility. We use Deep Reinforcement Learning (DRL) algorithms to address the challenges of the dynamic environment optimization problem.
% 为什么使用Diffusion Model 作为 Policy
However, popular RL algorithms select actions corresponding to different reward distributions by learning an optimal policy. As a result, these algorithms often tend to yield unimodal policy distributions, which may lead to convergence to local optima and hinder adaptation to environmental changes and complex scenarios \cite{yang2023policy}. Fortunately, the diffusion model specializes in capturing complex probability distributions \cite{yang2023diffusion}. When used as a policy network in the DRL algorithm, the diffusion model facilitates a more favorable trade-off between exploration and exploitation. Furthermore, considering that the state of the environment must take into account the characteristics of tasks, including their type and required resources, as well as the attributes of different edge servers, such as the attributes of deployed AIGC models and computational resources, the state vector is designed as a high-dimensional vector. We introduce the attention mechanism to capture the relationships between tasks and different edge servers in the high-dimensional vector. The main contributions are summarized as follows:
% 贡献
\begin{itemize}
    \item We propose a scenario for the widespread use of AIGC services under the condition of multi-type tasks supported by edge computing. 
    \item  Given the complex and dynamic nature of the wireless network, we introduce the diffusion model as a policy function within the DRL algorithm to capture complex probabilistic distribution.
    \item  We employed an attention mechanism to capture the long-range dependencies between task and edge server attributes in the state vector and proposed the ADSAC algorithm. Comparative experiments with currently popular DRL algorithms have demonstrated the effectiveness of our ADSAC algorithm.
\end{itemize}

\section{Related Works}
% AIGC服务、ASP选择方法
The ability of AIGC services to generate content both powerfully and efficiently has led to their gradual integration into various aspects of daily life. AIGC is currently advancing by taking advantage of wireless network developments and applying edge computing technologies to facilitate its widespread adoption. Du. and Li. \textit{et al.} have stated that different AIGC models produce different user utilities, and it is necessary to design an optimization algorithm for selecting appropriate AIGC models to improve user utilities \cite{du2024diffusion}. They considered the resource constraints of different AIGC service providers and proposed a diffusion-based RL algorithm to refine AIGC model selection. Furthermore, the conditions of wireless network communication were taken into account in the scenario of edge-supported AIGC services \cite{du2023user}, where computational and communication resources together act as constraints in the optimization process to maximize user utility.
% 与上述研究不同的是，我们考虑了多类型任务
Building on existing studies, our work introduced an additional consideration of the adaptability between AIGC models and user task types. It is based on the fact that an AIGC model can provide varying utilities across different task types, and different AIGC models possess distinct advantages for various specific task types. This fact leads us to include task type as a variable in the formulation of the user utility function and as one of the conditions in the optimization problem for AIGC model selection.

\section{Problem Formulation}

% 问题基于三个事实：用户任务多样、不同AIGC models对同一类型任务的用户效用不同、计算资源限制
The challenges of the AIGC service selection problem derive from three facts. First, user tasks are diverse and can be classified into multiple types, including architecture, portrait, and landscape. Second, different AIGC models provide distinct user utilities for the same type of task. Third, the computational resources of the edge servers hosting the AIGC models are constrained.
% 目标：用户效用最大化
Our work aims to maximize the user's utility by selecting appropriate AIGC models for different user tasks. 

% 问题建模
The scenario of our problem is edge-driven AIGC services with $I$ edge servers, and we assume that $J_i$ AIGC models are deployed on the $i^\text{th}$ edge server, where the $j^\text{th}$ AIGC model deployed on the $i^\text{th}$ edge server is denoted as $a_{ij}$.
% 问题建模为马尔可夫决策过程，由于AIGC任务没有ground truth
The problem can be formulated as a Markov Decision Process (MDP) since the AIGC task lacks a ground truth which is the generated image that maximizes its user utility. Details about the MDP formulation are provided in Section \ref{s5}.
% 任务到达建模
Each episode in MDP is independent, with an assumed duration of $\eta$. The Poisson process is appropriate for modeling the arrival of user tasks \cite{du2024diffusion}, thus the total number $N$ of the user tasks that require AIGC services is calculated by sampling the Poisson probability distribution with the arrival rate $\lambda$ in each episode. The type of the $n^\text{th}$ task $t_n$ is $K(t_n)$.

% 用户效用建模
The user utility for the task $t_n$ by the AIGC model $a_{ij}$ can be modeled as
\begin{equation}
U_{a_{ij}}(t_n)=U_{a_{ij}}(F_{a_{ij}}(t_n,R_{t_n}),K(t_n)),\label{eq2}
\end{equation}
where $F_{a_{ij}}(\cdot)$ is the forward function of the AIGC model and $R_{t_n}$ is the resource requested by the task. To avoid the factor of contingency, a baseline $\bar{U}_{a_{ij}}(F_{a_{ij}}(\cdot), K(t_n))$ is first established for the user utility resulting from the processing task of the AIGC model $a_{ij}$. We utilize a general image quality assessment metric with experimental experience and choose a version based on the Blind/Referenceless Image Spatial Quality Evaluator because it correlates with human perception\cite{du2023enabling}. Furthermore, they showed that the quality $Q_{a_{ij}}(R_{t_n})$ of the content generated by the AIGC model $a_{ij}$ is approximately proportional to the required resources $R_{t_n}$ (number of denoising steps in the diffusion model). By combining the baseline for tasks of a specific type generated by the AIGC model and the practical perceptual quality of generated content for a concrete task, we obtained the user utility function as
% We obtained the user utility function of selecting AIGC model $a_i$ for task $t_j$ as
\begin{equation}
U_{a_{ij}}(t_n) =\beta \cdot \bar{U}_{a_{ij}}(F_{a_{ij}}(\cdot), K(t_n)) + Q_{a_{ij}}(R_{t_n}),\label{eq5}
\end{equation}
where $\beta$ is the hyperparameter that balances these two metrics and is a trade-off between the basic capabilities against the real-time performance of the AIGC model.

% 资源限制
Resource constraints also require consideration in the problem. 
The sum of resources consumed by all the AIGC models deployed on the $i^\text{th}$ edge server should be less than the resource $R_i$ that can be provided by this edge server in the practical scenario, or else it may cause a crash. 
% 问题建模
Under this problem, each arrival task $t_n$ is mapped to an AIGC service selection scheme $\mathcal{A}_{t_n}$. The problem is formulated as
\begin{align}
\max_{\mathcal{A}} \quad & \mathcal{U} = \sum_{n=1}^{N} U_{a_{ij}}(F_{a_{ij}}(t_n,R_{t_n}),K(t_n)), \label{eq6} \\
\text {s.t.} \quad & a_{ij} = \mathcal{A}_{t_n}, \label{eq7} \\ 
                   & R_{t_n} + \sum_{j=1}^{J_i} \sum_{n=1}^{N_{a_{ij}}} R_{t^{\prime}_n} \leqslant R_i \quad (\forall i \in I), \label{eq8} \\ 
                   & i=1, \ldots I, \; j=1, \ldots J, \; \text{and} \; n=1, \ldots N. \label{eq9}
\end{align}
The available resource constraint of edge servers is reflected in constraint \eqref{eq8}, where $N_{a_{ij}}$ denotes the number of tasks being serviced by AIGC model $a_{ij}$ deployed on the $i^\text{th}$ edge server at the time of the task $t_n$ allocation. Therefore, we aim to design an algorithm that finds the optimal solution $\mathcal{A}$ to maximize the overall user utility $\mathcal{U}$ derived from the generated content across all task counterparts.

% ASP selection problem under multi-types of user tasks
\section{Attention-based Diffusion Soft Actor-Critic for AIGC Service Selection} \label{s5}
\subsection{Denoising Diffusion Probabilistic Model}
% 引入
Considering the diffusion model can use state information to denoise stochastic noise to capture complex probability distributions, we use it as the policy network for the online RL algorithm to generate a vector of AIGC service selection schemes. We choose the Denoising Diffusion Probabilistic Model (DDPM) \cite{ho2020denoising}, a typical diffusion model.

% Forward Process
The DDPM exhibits two trajectories in opposing directions. One of the trajectories executes Gaussian noise addition to the expert scheme $x_0 \sim q(x_0)$ during $T$ timesteps so that $x_0$ gradually becomes Gaussian noise $x_T$. The process can be described as \cite{ho2020denoising}
\begin{equation}
q(x_{1:T} \mid x_0) = \prod_{t=1}^T \mathcal{N}(x_t ; \sqrt{1-\beta_t}x_{t-1}, \beta_t \mathbf{I}),\label{eq10}
\end{equation}
where $q(x_{1:T} \mid x_0)$ is called the forward process or diffusion process, and $\beta_t$ can be held constant as hyperparameters to control the variance of the Gaussian noise at timestep $t$. Mathematically expanding and combining each step in \eqref{eq10} with $\alpha_t = 1-\beta_t$ and $\bar{\alpha}_t=\prod_{s=1}^t \alpha_s$.

The other trajectory involves $T$ timestep denoising operations on the Gaussian noise $x_T$ obtained from sampling $\mathcal{N}(0,\mathbf{I})$ to generate a scheme $\hat{x}_0$, and can be described as \cite{ho2020denoising}
\begin{equation}
p_\theta(x_{0:T}) = p(x_T)\prod_{t=1}^T \mathcal{N}(x_{t-1} ; \mu_\theta(x_t, t, s), \Sigma_\theta),\label{eq12}
\end{equation}
where $p_\theta(x_{0:T})$ is called the reverse process or denoising process, $s$ is the state information, and $\theta$ is the model parameter. The goal of optimization is to approximate the means and variances of $q(x_{t-1} \mid x_t,x_0)$ and $p_\theta(x_{t-1} \mid x_{t})$. Thereby, We obtain the mean of the reverse process as \cite{ho2020denoising}
\begin{equation}
\mu_\theta(x_t, t, s) = \frac{1}{\sqrt{\alpha_t}} (x_t -\frac{1-\alpha_t}{\sqrt{1-\bar{\alpha}_t}} \epsilon_\theta(x_t, t, s)),\label{eq16}
\end{equation}
and
\begin{equation}
\Sigma_\theta = \frac{1-\bar{\alpha}_{t-1}}{1-\bar{\alpha}_t} \beta_t \mathbf{I}.\label{eq15}
\end{equation}
where $\epsilon_\theta(\cdot)$ is a noise prediction network parameterized by $\theta$. Thereby, $x_{t-1} \sim p_\theta(x_{t-1} \mid x_{t})$ resulting from denoising $x_t$ can be computed by
\begin{equation}
x_{t-1} = \frac{1}{\sqrt{\alpha_t}} (x_t -\frac{1-\alpha_t}{\sqrt{1-\bar{\alpha}_t}} \epsilon_\theta(x_t, t, s)) + \sqrt{\beta_t} \mathbf{z},\label{eq17}
\end{equation}
where $\mathbf{z} \sim \mathcal{N}(0,\mathbf{I})$.

\begin{figure}[!t]
    \centering
    \includegraphics[width=\linewidth]{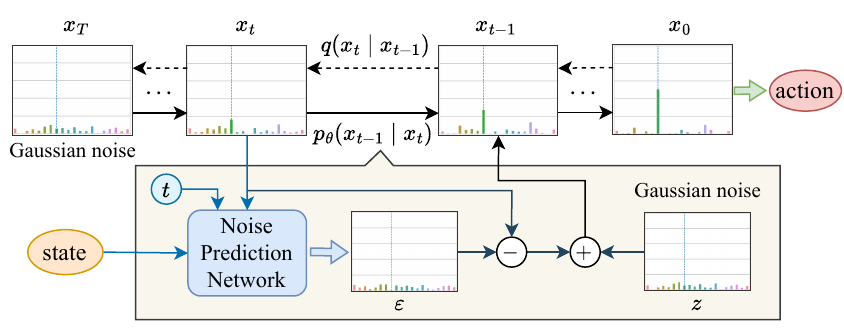}
    \caption{Structure of DDPM as policy in reinforcement learning algorithm.}
    \label{fig2}
\end{figure}

For the AIGC model selection problem, establishing expert scheme sets is time-consuming and not always possible, thus we disregard the forward process. Instead, we focus on the reverse process of the DDPM which is used as the policy network $\pi_\theta(s)$, shown in Fig.~\ref{fig2}. The policy network takes the environmental observation state as input and predicts the noise at each timestep. Subsequently, the policy network generates the action probability vector by denoising using \eqref{eq17} and obtains the AIGC model selection scheme.

\subsection{Attention-based Diffusion Soft Actor-Critic}
% MDP建模
To mitigate the risk of edge servers crashing caused by overloading and optimize overall user utility, the problem of selecting the optimal AIGC model service for arrival tasks can be effectively modeled and solved using MDP. MDP is encapsulated by the quadruple $\langle \mathcal{S}, \mathcal{A}, r(\cdot), P(\cdot) \rangle$, where $\mathcal{S}$ and $\mathcal{A}$ represent the set defining the state space and action space, respectively. The function $r(\cdot)$ denotes the reward function which yields a numeric reward for each state-action pair, and $P(\cdot)$ signifies the state transition probability function describing the likelihood of transitioning between states for given actions. This quadruple framework provides a comprehensive model for sequential decision-making in stochastic environments and is described as follows.

\textbf{State Space.} Within the state space $\mathcal{S}$, each state $s$ is conceptualized as a vector that integrates the requisite information for action determination. This integration encompasses characteristics of the current task and critical server-related information. In terms of task, $s$ incorporates encoding of task type $K(t_n)$, the computational resources demanded $R_{t_n}$, and the completion timeframe $O_{t_n}$. In terms of server, $s$ accounts for the aggregate resources $R$, the proportion of remaining resources $\tilde{R}$, and the quantity of deployed AIGC models across each server. Consequently, the state $s$ is represented as $[K(t_n), R_{t_n}, O_{t_n}, R_1, \tilde{R}_1, J_1, \dots, R_I, \tilde{R}_I, J_I]$.

\textbf{Action Space.} The action space $\mathcal{A}$ is defined as an integer domain, representing the sequential identifier of an AIGC model (according to putting AIGC models owned by each server in order), allocated to the current task.

\textbf{Reward Function.} The reward function depends on the current state and action obtained from the policy function. Considering the resource limitations of edge servers, the reward function is delineated into two distinct cases. In the case that the server on which the designated AIGC model is hosted possesses adequate resources to execute the task, the reward function is the user utility function, denoted as
\begin{equation}
r(s, a)=U_a(F_a(t_n, R_{t_n}), K(t_n)).\label{eq18}
\end{equation}
Conversely, in the case that resource inadequacy precipitates server failure, the reward function is designed as a penalty, denoted as $r(s, a) = -r_p(s, a)$. Specifically, $r_p(s, a)$ is determined based on a fixed penalty $p$ and the progress $G(t^\prime_n)$ of task being served by the crashed server \cite{du2024diffusion}, as
\begin{equation}
r_p(s, a) = p \cdot (1 + \sum_{j=1}^{J_i} \sum_{n=1}^{N_{a_{ij}}} (1-G(t^\prime_n))).\label{eq19}
\end{equation}

% \begin{figure}[!t]
%     \centering
%     \includegraphics[width=\linewidth]{figs/fig3.pdf}
%     \caption{Architecture of the ADSAC algorithm.}
%     \label{fig3}
% \end{figure}

\begin{table*}[!t]
    \caption{Performance Comparisons of ADSAC with Popular Value-based Reinforcement Learning Algorithms, Policy-based Reinforcement Learning Algorithms, and Heuristic Algorithms.}\label{tab1}
    \centering
    % \normalsize
    \small
    \begin{tabular}{cccccc}
        \toprule
        \multicolumn{2}{c}{Method}        & Train Reward      & Test Reward       & Crashed Rate    & Lost Utility \\ 
        \midrule
        \multirow{4}{*}{Heuristic Methods}& Random      & 471.88            & 514.089           & 36.64\%      & 423 \\
        & Round Robin   & 640.449           & 614.593           & 31.16\%         & 370          \\
        & Crash Avoid   & 1298.325          & 1239.065          & 0.41\%          & 14           \\
        & Prophet       & 1969.11           & 1927.026          & 0.33\%          & 2            \\ 
        \midrule
        \multirow{6}{*}{DRL Methods}& DRQN        & 1589.674          & 1489.964        & 13.26\%         & 235 \\
        & Rainbow       & 1711.303          & 1676.008          & 4.07\%          & 72           \\
        & PPO           & 1210.395          & 952.774           & 42.82\%         & 737          \\
        & SAC           & 1762.472          & 1641.758          & 2.93\%          & 40           \\ 
        & DSAC        & 1796.227          & 1715.917          & 1.13\%          & 12           \\
        & ADSAC       & \textbf{1843.865} & \textbf{1779.501} & \textbf{0.47\%} & \textbf{5}   \\ \bottomrule
    \end{tabular}
\end{table*}

% We proposed an attention-based diffusion model leveraging self-attention in the noise prediction network of the diffusion model to capture critical dependencies in input vectors, as shown in Fig.~\ref{fig3}.

We proposed an attention-based diffusion model leveraging self-attention in the noise prediction network of the diffusion model to capture critical dependencies in input vectors. Specifically, the state $s_{t^\prime}$ at the $t^\prime$ step and the generative vector $x_t$ at the $t$ timestep are processed through multi-layer perceptrons separately for initial feature extraction. The model incorporates sinusoidal positional embedding \cite{vaswani2017attention} to encode the timestep $t$, enriching temporal information in the reverse process. After vector extraction and positional embedding, these features are concatenated to form a feature vector $f$. The composite feature vector $f$ is fed to multi-layer perceptrons with parameters $W^q$, $W^k$, and $W^v$ to obtain the query, key, and value of the attention mechanism, respectively. We adopted the scaled dot-product attention mechanism \cite{vaswani2017attention}. The output of the self-attention layer is mapped to the predicted noise using a multi-layer perceptron as a decoder, which is utilized to denoise $x_t$ for $x_{t-1}$.

Our proposed ADSAC algorithm incorporates the attention-based diffusion model as the policy network of the SAC algorithm. SAC algorithm is a policy-based RL, recognizing its superior sample efficiency by offline policy learning and its enhanced exploratory capabilities by the maximum entropy RL \cite{haarnoja2018soft}. 
% The architecture of the ADSAC algorithm is shown in Fig.~\ref{fig3}, 
The ADSAC algorithm includes a pair of critic networks $Q_{\omega_1}(s, a)$ and $Q_{\omega_2}(s, a)$, corresponding target networks $Q_{\omega^-_1}(s, a)$ and $Q_{\omega^-_2}(s, a)$, and a policy network $\pi_\theta(s)$. In this process, actions are selected iteratively based on the policy network informed by the current state of the environment. Offline policy learning stores the transition $(s_{t^\prime}, a_{t^\prime}, r_{t^\prime}, s_{t^\prime+1})$, denoted by $\zeta_{t^\prime}$, in an experience replay buffer to reduce temporal correlation and improve sample efficiency.

Offline policy learning algorithm updates models by sampling a batch $\mathbf{B}$ of transitions from the experience replay buffer. Critic networks estimate the action value based on $s_{t^\prime}$ and $a_{t^\prime}$ and are updated utilizing the temporal difference(TD) learning algorithm with the loss function as
\begin{equation}
L_Q(\omega) = \frac{1}{| \mathbf{B} |} \sum_{\zeta_{t^\prime} \in \mathbf{B}} [Q_\omega(s_{t^\prime}, a_{t^\prime}) - (r_{t^\prime} + \gamma V_{\omega^-}(s_{t^\prime+1}))]^2,\label{eq22}
\end{equation}
where $\gamma$ is the discount rate for the return. The loss function used to update the policy network is
\begin{equation}
L_\pi(\theta) = \frac{1}{| \mathbf{B} |} \sum_{\zeta_{t^\prime} \in \mathbf{B}} (\min_{i=1,2} Q_{\omega_i}(s_{t^\prime}, a_{t^\prime})) + \alpha H(\pi_\theta(\cdot \mid s_{t^\prime})).\label{eq25}
\end{equation}
The state value function is calculated as 
\begin{equation}
V_{\omega^-}(s_{t^\prime}) = \min_{i=1,2} Q_{\omega^-_i}(s_{t^\prime}, a_{t^\prime}) + \alpha H(\pi_\theta(\cdot \mid s_{t^\prime})),\label{eq26}
\end{equation}
where the minimum output between two critic networks is selected to mitigate the problem of overestimation. Therefore, the policy network and critic networks are updated as
\begin{equation}
\begin{split}
& \omega_i \leftarrow \omega_i - \delta_Q \nabla_{\omega_i} L_Q(\omega_i), \quad i \in {1,2}, \\
& \theta \leftarrow \theta - \delta_\pi \nabla_\theta L_\pi(\theta), \\
& \omega_i^- \leftarrow \tau \omega_i + (1-\tau)\omega_i^-, \quad i \in {1,2}, \label{eq27}
\end{split}
\end{equation}
where $\delta_Q$ and $\delta_\pi$ represent the learning rates of the critic network and the policy network respectively, and $\tau$ denotes the soft update coefficient for the target critic network. Our proposed ADSAC algorithm iteratively collects transitions for storage in the experience replay buffer, and samples mini-batches from this buffer to perform model updates, until convergence is achieved with maximization of overall user utility.

\begin{figure*}[!t]
    \centering
    \subfloat[]{\includegraphics[width=0.32\textwidth]{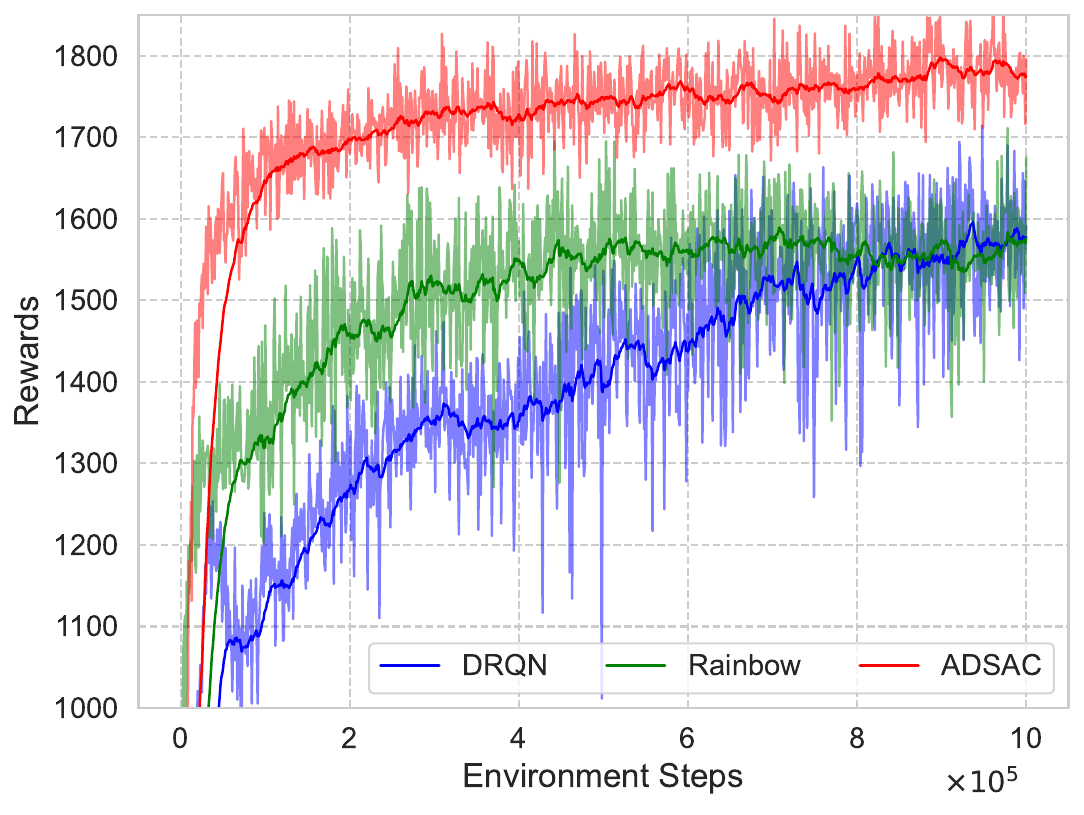}
        % \caption{}
        \label{fig4:1}}
    \hfill
    \subfloat[]{\includegraphics[width=0.32\textwidth]{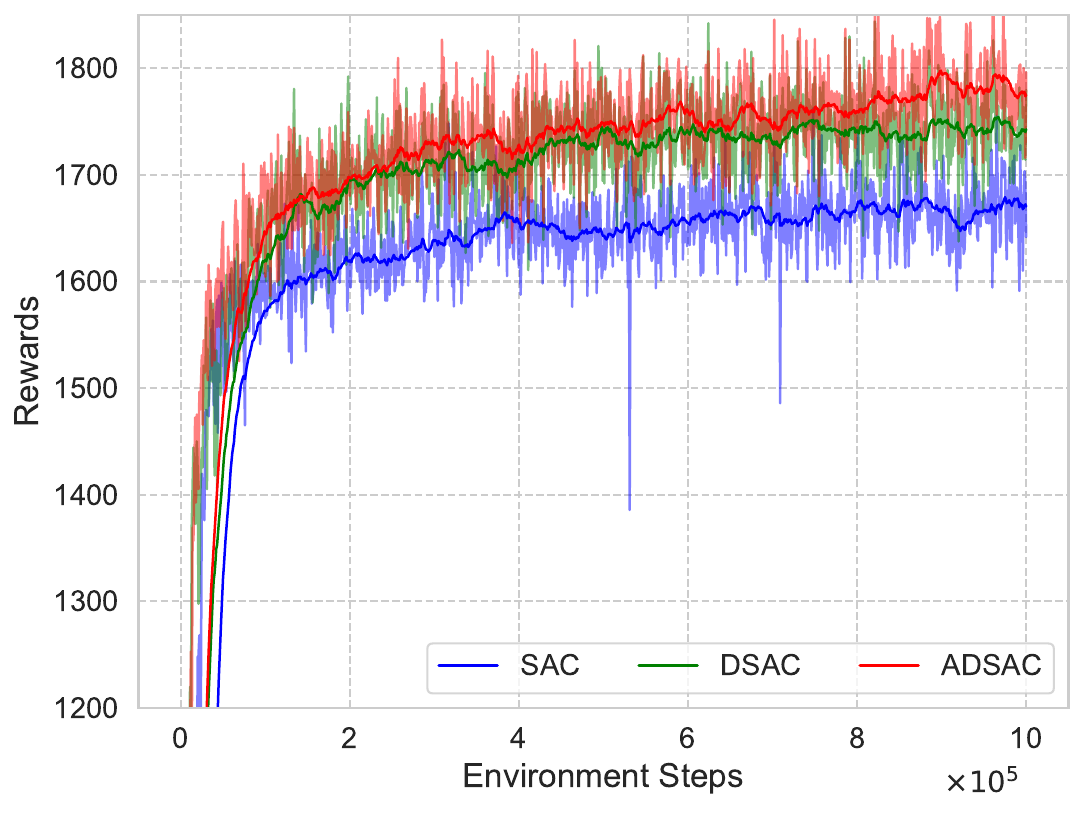}
        % \caption{}
        \label{fig4:2}}
    \hfill
    \subfloat[]{\includegraphics[width=0.32\textwidth]{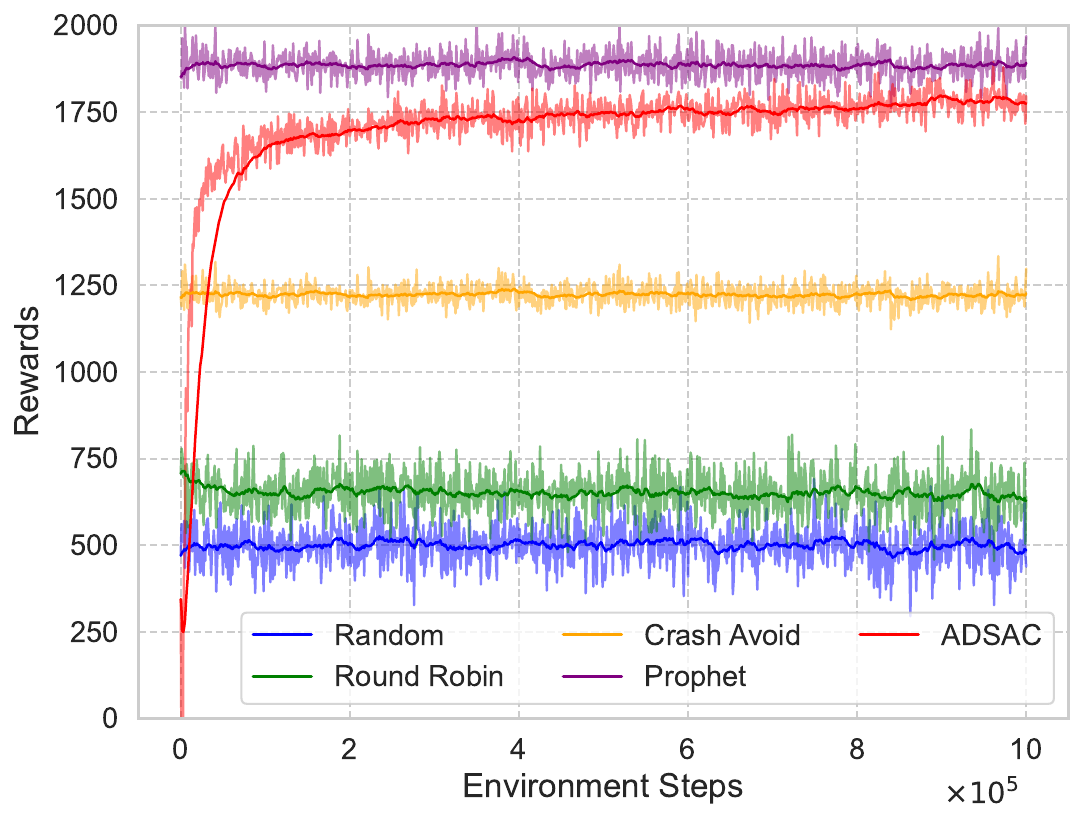}
        % \caption{}
        \label{fig4:3}}
    \caption{Comparison of test reward curves of the ADSAC algorithm with popular value-based reinforcement learning algorithms, policy-based reinforcement learning algorithms, and heuristic algorithms, respectively.}
    \label{fig4}
\end{figure*}

\section{Experiments}
\subsection{Experimental Setup}
The simulation experiments consider a scenario with 5 edge servers, each with 4 AIGC models deployed on it. Existing studies have shown that energy consumption is proportional to the number of denoising steps for diffusion model-based generation services \cite{du2024diffusion}. This relationship allows for a simplification of resource considerations to merely account for the denoising steps. The total resource capacity allocated to each edge server is randomized, falling within a range from 1500 to 3000. The simulations consider four distinct types of tasks, with the denoising steps required for each task randomly determined, varying between 100 and 250 steps. Training of 1000 epochs in total, 1000 transitions are collated for each epoch. The discount factor $\gamma$ for future returns is set at 0.95. Each simulation episode spans a duration $\eta$ of $10^6$ time units, with task arrivals $\lambda$ as 0.0015 in the Poisson distribution. 

For the ADSAC algorithm, the target network is updated using a soft update coefficient $\tau = 0.005$, and it employs an entropy regularization coefficient of $\alpha = 0.05$ to encourage exploration. The learning rate for the action network $\delta_\pi$ is set to $10^{-4}$ and critic networks are set a higher learning rate $\delta_Q$ of $10^{-3}$. The action network uses a diffusion model with a total of 5 time steps. Regarding the network architecture, the dimensions of output features for the linear networks which are for extracting $x_t$ and the state are 32 and 64, respectively. Furthermore, the feature vector derived from the self-attention mechanism is processed through a multilayer perceptron, which has a hidden layer size of 256, to predict the noise.

\subsection{Performance and Analysis}
The reward variations in each episode during the training of algorithms as the learned environment steps increase are shown in Fig.~\ref{fig4}. We benchmark the performance of the ADSAC algorithm against popular value-based RL algorithms, policy-based RL algorithms, and heuristic approaches. In the comparison with value-based RL algorithms,  we observe the sluggish convergence for the DRQN, and the demonstrates superior convergence speed and performance of ADSAC algorithm. In the comparison with policy-based RL algorithms, both the DSAC and ADSAC algorithms, which leverage the diffusion model, surpass the performance of the SAC algorithm that relies on a multilayer perceptron. The DSAC algorithm has the same settings as the ADSAC algorithm except that it is not using self-attention. Notably, the incorporation of the self-attention mechanism renders the ADSAC algorithm the most effective among them. The PPO algorithm is omitted due to its subpar performance. Compared with heuristic algorithms, the ADSAC algorithm approximates the prophet solution closely, indicative of its strategic decision-making to maximize reward at each step. The crash-avoid strategy exhibits limited effectiveness, primarily because it focuses solely on preventing system crashes without considering the compatibility of the AIGC models with the types of user tasks. 

A detailed comparison of the numerical performance metrics for each algorithm is presented in Table \ref{tab1}. The crash rate is defined as the ratio of crashed tasks that failed to be completed successfully to the total tasks, and lost utility represents the cumulative lost utility due to the crashed tasks. Excluding the impractical prophet solution and crash-avoid strategy, our ADSAC algorithm emerges as the superior performer across all evaluated metrics. Furthermore, the performances of DRL algorithms that incorporate the Diffusion model over their counterparts underscores the significant contribution of the diffusion model to enhancing the generation and optimization of AIGC model selection strategies.

\begin{figure}[!t]
    \centering
    \includegraphics[width=0.8\linewidth]{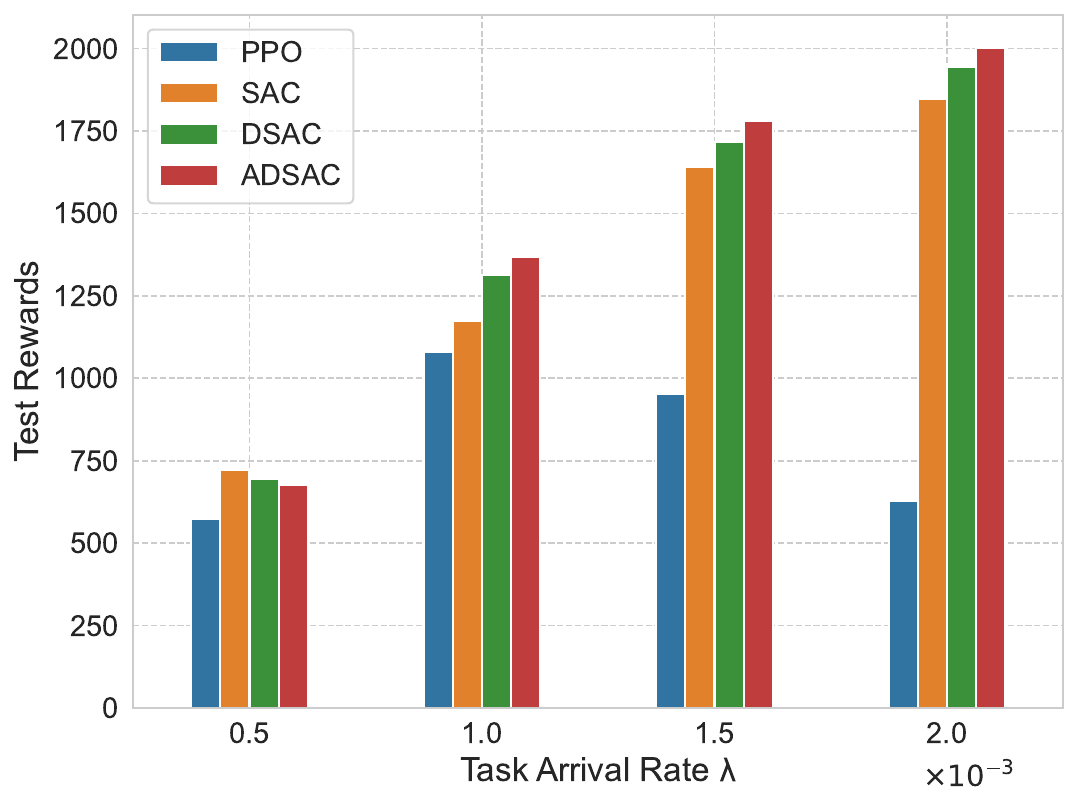}
    \caption{Test rewards over different task arrival rates.}
    \label{fig5}
\end{figure}

We conducted a further analysis of the performance of algorithms across environments characterized by varying $\lambda$, which represent different levels of task crowding, as illustrated in Fig.~\ref{fig5}. SAC, DSAC, and ADSAC algorithms show resilience to task crowding scenarios. However, the PPO algorithm exhibited a notable sensitivity to task congestion, struggling to manage increased task volumes effectively. The ADSAC and DSAC algorithms showed superior handling capabilities in environments of task crowding when compared to SAC and PPO, with ADSAC consistently outperforming the others. The ADSAC algorithm notably maintains the highest test rewards at arrival rates of 0.001, 0.0015, and 0.002, making it the most efficient solution of all the evaluated algorithms.

\section{Conclusion}
This paper explores the compatibility between multi-type user tasks and AIGC models in the deployment of AIGC models on edge servers. We formulated the user utility function to reflect the inherent generative capabilities across different tasks and the specific real-time performance of the AIGC model. We employed the DRL algorithm to optimize overall user utility in the AIGC model selection problem. Given the distinguished performance in diverse scenarios and robust feature extraction capability of the diffusion model, it served as the policy network in our DRL framework. Furthermore, to address the contextualized correlation in long state vectors, we integrated the attention mechanism with the diffusion model and proposed the ADSAC algorithm. Comparative experiments against popular DRL and heuristic approaches demonstrated that the ADSAC algorithm not only excels in preventing edge server crashes but also selects the AIGC model suitable for the task type adeptly.

% \section*{Acknowledgment}

% The preferred spelling of the word ``acknowledgment'' in America is without 
% an ``e'' after the ``g''. Avoid the stilted expression ``one of us (R. B. 
% G.) thanks $\ldots$''. Instead, try ``R. B. G. thanks$\ldots$''. Put sponsor 
% acknowledgments in the unnumbered footnote on the first page.

\bibliographystyle{IEEEtran}
\bibliography{myref}
% \end{thebibliography}

\vspace{12pt}
% \color{red}
% IEEE conference templates contain guidance text for composing and formatting conference papers. Please ensure that all template text is removed from your conference paper prior to submission to the conference. Failure to remove the template text from your paper may result in your paper not being published.

\end{document}